\documentclass{article}
\usepackage{ltwol-doug}
\usepackage{epsf}
\usepackage{fleqn}

\def\be{\begin{equation}}
\def\ee{\end{equation}}
\def\bea{\begin{eqnarray}}
\def\eea{\end{eqnarray}}

\begin{document}

\title{HIGH ENERGY COSMIC NEUTRINOS}



\author{STEVEN W. BARWICK}

\address{Dept. of Physics and Astronomy, University of California-Irvine, 
Irvine, CA, USA}


\twocolumn[\maketitle\abstracts{
While the general principles of high-energy neutrino detection have
been understood for many years, the deep, remote geographical locations of
suitable detector sites have challenged the ingenuity of experimentalists,
who have confronted unusual deployment, calibration, and robustness issues.
Two high energy neutrino programs are now operating (Baikal and
AMANDA), with the expectation of ushering in an era of multi-messenger
 astronomy,
and two Mediterranean programs have made impressive progress. The detectors
are optimized to detect neutrinos with energies of the order of 
1-10 TeV, 
although they are capable of detecting neutrinos with energies of 
tens of MeV to greater than PeV.  This paper 
outlines the interdisciplinary scientific agenda, which span the fields of
 astronomy, particle
physics, and cosmic ray physics, and describes ongoing worldwide experimental 
programs to realize these goals.}]

\section{Introduction}
The high energy frontier has traditionally led to dramatic 
breakthroughs in our understanding of nature.  High energy neutrino detectors
are designed to probe of some of the most violent and energetic 
phenomena in the
Universe.  Neutrinos born in the hearts of these phenomena provide a unique
view of how nature accelerates particles and clarify the role of hadrons in 
the astrophysical milieu.  Once produced, neutrinos are unaffected by
intervening matter or photons. Being uncharged, they propagate through
the universe undisturbed by magnetic fields. Given the current uncertainty
in the location of the sources of extremely energetic cosmic rays, the
neutrino messenger may be the only route to clear identification.

Theorists have identified a variety of potential sites of high energy neutrino
production, and several extensive reviews of this topic have appeared 
recently in the literature \cite{pr,TASI98,Proth99}.
For example, Protheroe has summarized the astrophysical
predictions of diffuse neutrino intensities between 1 TeV and the GUT scale.
Recently, Gamma Ray Bursts (GRBs)have occupied the theoretical spotlight with
the discovery that they are distant extragalactic phenomena and 
therefore the most energetic events
observed in the Universe.   Waxman and Bahcall \cite{WB97} have 
argued that
GRBs are {\it the} sources of the extremely high energy (EHE) cosmic rays and
prodigious
sources of high energy neutrinos.  The predicted flux is tied to the
measured power density of EHE cosmic rays, which also 
has been used to constrain the
neutrino flux in proton blazer models of AGN \cite{WB99}. Though this
procedure is still generating significant debate \cite{Mannheim99,TASI98}, 
there is no doubt that models should not over-produce cosmic rays.

Just as multi-wavelength studies have provided unparalleled
insight on many astronomical sources, multi-messenger studies by neutrino,
gamma ray, and gravity wave detectors may be the Rosetta stone of cosmic
accelerators. For example, the AMANDA neutrino facility, located at the 
South Pole, contemporaneously observes the same sky as new, powerful gamma 
ray telescopes in the northern hemisphere. Coincidence experiments can
also be contemplated with space-based gamma ray observatories and 
gravitational wave detectors such as LIGO or VIRGO.  At the very highest
energies, charged cosmic rays are expected to deviate only slightly from
line-of-sight trajectories.  Should the Hi Res and Auger Observatories  
identify sources of extremely energetic particles, then concurrent 
observations by neutrino telescopes can provide additional information on
the local environment of the accelerator.

In all models of particle acceleration
to extreme energies, the flux decreases with energy. The decreasing flux
is not fully compensated by an increasing  sensitivity of the neutrino
detector.  This leads to a strategy that concentrates on somewhat 
lower energy neutrinos, acting as surrogates for the extremely energetic,
but far rarer, cosmic rays.

The essential characteristics of a neutrino telescope have been known for
more than two decades \cite{history60,berez77}. Markov suggested in 1960
that
the ocean would be a suitable site for constructing a large neutrino detector
based on the detection of Cherenkov light, and most important features were
discussed and specified during a series of workshops devoted to developing 
the DUMAND concept. Halzen and Learned \cite{icehistory} introduced a 
twist on the general scheme by promoting polar ice as suitable
medium. Until recently, workable implementations of these sensible ideas 
have been thwarted by unusual technical and logistical challenges associated
with the remote deployment of hardware in media that differ from ordinary
 purified water in severalimportant details.
All current architectures for high energy neutrino facilities 
bury a sparse array of optical sensors
within deep ice, ocean or lake waters. The optical sensors respond
to the UV dominated cherenkov radiation emitted by neutrino-induced muons
or neutrino-induced hadronic and electromagnetic cascades.  Large 
detector volumes are 
required because the predicted flux of cosmic neutrinos and the known 
interaction probabilities at the energies of interest are relatively small. 
The detection probability, defined as the ratio between 
the range of the muon to the interaction mean free path of the neutrino, 
is only $10^{-6}$ for a $\nu_{\mu}$ with an energy of 1 TeV. Moreover,
the rare signal events must be extracted from a large flux of atmospheric
muon background.
For example, at sea level the number of background muons per 
unit area exceeds the expected neutrino-induced
muon signal by $\sim\! 10^{11}$, so neutrino detectors are constructed 
at large depths to reduce 
this unwanted signal.  Even at depths of 2 km of water equivalent, 
down-going background exceeds 
predicted signal by a factor of $\sim 10^5$. The combination of large volume, 
large overburden, and desire to minimize material costs leaves 
experimentalists with few options other than to construct a detector within a 
remote, naturally occurring, transparent medium such as ice or water 
(no excavated caves or mines are large enough). The formidable technical 
challenge of remote operation distinguishes high energy neutrino facilities 
from existing solar and accelerator-based neutrino detectors. It is one
factor which has spurred the development of surface detectors (eg., GRANDE
and HANUL \cite{HANUL})despite the daunting background difficulties.

Cherenkov techniques are now well understood and illustrated below 
(see Fig. 1). 
\begin{figure}[h] 
\centering
\epsfxsize= 8cm\epsffile{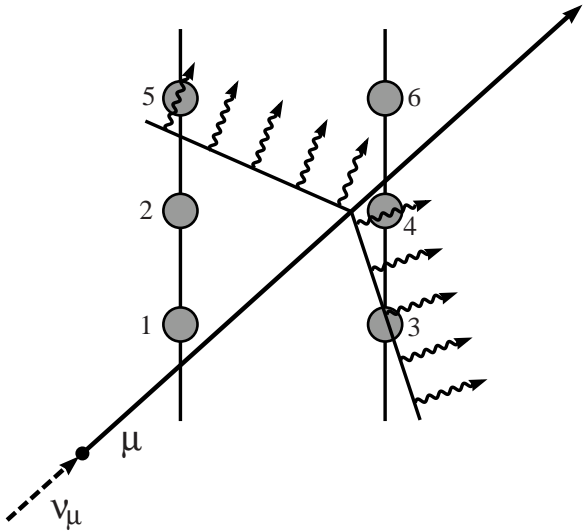}

\caption{Muon trajectories can be reconstructed by timing the passage of
the Cherenkov wavefront. The numbers indicate the time sequence of the
arrival of the photons.}
\end{figure}
A high energy neutrino can be detected only if it
converts to a charged lepton such as a muon.  Astronomy
is possible because the muon direction is aligned with the
incident neutrino to within a degree,  if the energy 
is greater than 1 TeV.  The angular correlation between charged lepton and 
neutrino improves as the $1/\sqrt{E}$,
so eventually multiple coulomb scattering becomes the dominate factor
in the angular resolution. Conceivably, neutrino directions can be 
located to $\sim\! 0.1^{\circ}$ in some designs \cite{ANTARES98}. Source
localization can be improved by the detection of multiple events, but
unless the event rate is unexpectedly large, the angular resolution is
not competative with conventional astronomy.  Therefore sources must be 
identified statistically - by searching for a class of objects that 
lie within the angular error boxes.
Confidence will be bolstered if theoretical models of
that class of objects are consistent with high energy neutrino production.
The relatively limited number of potential sites of high energy
neutrino production suggests that source confusion is unlikely to be a problem.

The muon is detected by distributing photon sensors 
(large diameter photomultiplier tubes - PMTs) over the 
largest possible volume of transparent medium and recording the arrival times
and intensity of the Cherenkov wavefront.  Accurate reconstruction relies
on actively tracking events over linear dimensions exceeding tens of meters
and measuring the arrival of the Cherenkov wavefront to tens of nanoseconds
or better.
Geometries of the arrays are optimized according to the optical properties 
of the
detector media -- those media that generate less precision in the 
arrival time of 
the Cherenkov wavefront can be compensated by larger detectors with greater
average pathlength. Instrumented volume can be increased by utilizing
a medium with a large optical attenuation length. Naturally, volumes increase
with with additional sensors, so per unit costs become an important
design factor.

Muons from neutrino interactions are 
distinguished from the vastly more numerous atmospheric muons by direction;
upward-traveling muons (through the detector) can only originate from
nearby neutrino interactions.  The earth filters out all other known
particles.  Great care must 
be taken to reject the "down-going" atmospheric muons.  In practice,
muons are properly reconstructed if they traverse typically $\sim\!100m$ of
pathlength within the boundaries of array defined by the outermost 
strings, although dense arrays have demonstrated good reconstruction
with shorter tracks.    Complications arise from
the lack of fixed fiducial volume, the presence of events containing 
multiple muons, decaying muons in flight, and fluctuations in the
generation of Cherenkov photons resulting from high energy physics processes.  
Muon trajectories can pass near enough to trigger the array, but
too far outside the detector boundary for proper reconstruction.

Reconstruction is tied to specific assumptions about the event 
topology.  For example, the usual assumptions for a neutrino-induced
muon event are: 1) only one particle, 2) uniformly ionizing, 3)
travels at the speed of light, and 4) traverses the entire detector.
Deviation from these assumptions, such as stopping muons or decays in flight,
multiple-muon events, or muon-bremsstrahlung result in poorer reconstruction. 
Once the event is reconstructed, selection criteria must be established 
that reject events that are likely to be poorly reconstructed while 
maintaining good efficiency for signal events.

Neutrino-induced electrons produce electromagnetic cascades that generate 
very bright, localized bursts of Cherenkov photons.  While the
directional information is poor compared to muon tracks, the energy
resolution is far superior.  In media with moderate scattering, the
sensors nearest the cascade vertex provide the directional information,
while distant sensors sample from a expanding diffusive wavefront to
provide a calorimetric measurement. The spherical topology of the cascade 
events readily distinguish them from the most common atmospheric muon 
backgrounds.
Therefore, muon-induced bremsstrahlung and pair production become the dominant
background.  In this sense, the techniques of detecting muon-neutrinos
and electron neutrinos are complementary.  The good angular precision 
and superior sensitivity of muon detection is traded for improved energy 
resolution and lower background rates.
At energies above 1 TeV, the irreducible flux of atmospheric
$\nu_e$ is less than $\nu_{\mu}$ because fewer atmospheric muons
decay before reaching the detector as the muon energy increases.

As mentioned, the dominant source of background in high energy neutrino 
detectors
is downward  muon tracks generated by cosmic ray interactions in the 
atmosphere. This background can be avoided by constructing a detector at
$\geq$10 kmwe (kilometers of water equivalent) depths, but such depths are 
logistically impossible to attain. Rather, large volume
detectors are constructed at intermediate depths, and the background must be
removed by other methods.  In principle, the angular direction distinguishes 
astrophysical neutrino signals from the background of atmospheric muons - 
muons originating from below the horizon must originate from neutrino
interaction. However, errors in the reconstructed direction of muon
trajectory can result in misinterpreting down-going muons as upward
going muons.  For detector sites at depths between 1 and 4 kmwe,
and energy thresholds of $\sim\!10 GeV$,
the rate of down-going muons exceed potential signal rates by factors
of $10^3 - 10^5$ (assuming atmospheric neutrinos as signal).  
Therefore, an important design specification involves the rejection factor, R,  defined as $A_{eff}(signal)/A_{eff}(mis)$, where
 $A_{eff}(mis)=F_{m}\times A_{eff}(\mu_{atm})$, $A_{eff}(\mu_{atm})$
is the effective area for the detection of down-going muons, and $ F_m$ is the
fraction of down-going muons misidentified as upward going. The rejection
factor must be greater 
than $10^3$ for the best case conditions.   In the simplest description, 
$ F_m$ is a constant, but it may be treated as an angular dependent 
scattering probability $ P(\theta, \theta^{'})$ in more complex descriptions.
As the energy threshold of the detector is
increased to $\sim 10^{15}eV$, the ratio of downgoing atmospheric muons 
to expected
signal decreases, reaching unity in the vicinity of 1 PeV.  Since
the required level of rejection is less at higher energy thresholds,
event selection criteria can be optimized to achieve much larger effective
areas than could be achieved with larger rejection requirements. 
Detection methods with sufficient energy 
resolution to identify PeV events can be used to search the entire sky.  
Simulations show \cite{Porrata} that the energy of $\nu_{e}$-induced 
cascades may be measured with 
sufficient accuracy, assuming the vertex is contained within the volume 
of the array. 
The quoted values in the literature for effective detection area cause
 much confusion  because they are a
 function of lepton energy, zenith angle, and required rejection factor
which differs between physics objectives.  The effective volume becomes
useful when
the range of the muon is comparable to the largest dimension of the array.
 For muon detection at medium energies (and for all cascade events), the 
effective volume becomes
a convenient parameter of detector sensitivity, but it too depends on
energy and rejection factor. 

Atmospheric neutrinos form an irreducible background in the sense that they
cannot be differentiated from non-terrestrial neutrino signals on an event
by event basis.  Since the energy spectra and angular distributions of
atmospheric neutrinos are reasonably well known from measurement and 
calculation, statistical techniques 
using energy spectra, spatial and temporal correlations, etc. can confirm
or reject a hypothesis involving atmospheric neutrinos. 

The multifaceted scientific objectives of high energy neutrino telescopes 
are distributed across the fields of cosmic ray physics, astronomy, 
and particle physics. This diversity emphasizes the interdisciplinary 
potential of these
detectors. Two high energy neutrino programs are now operating 
(Baikal and AMANDA) and two Mediterranean programs have made 
impressive progress. Neutrino observatories
are optimized to detect neutrinos with energies of the order of 1-10 TeV, 
although they are capable of detecting neutrinos with energies of 
tens of MeV to greater than PeV. These detectors are distinguished by
the broad range in energy response.
 
\section{Science Goals}
 
 The scientific agenda is too broad to be covered within the limited
space of this paper, so I discuss only a few examples. Readers interested
in greater detail should consult the reviews referenced in the introduction.
The physics goals can be categorized according to the energy of the 
neutrino: low ($\sim\!10 MeV$), 
medium (10-100 GeV), and high ($\geq$1 TeV).  A 
transient burst of low energy neutrino emission from Supernova explosions
or Gamma Ray Bursts (GRBs) can be detected by summing the random noise 
signals from the photomultiplier
tubes in the optical modules within the array. A supernova burst would 
manifest itself as a statistically significant increase in the summed
signal due to the excess photons generated by the low energy neutrino
interactions. Sensitivity to transient events
is improved by embedding the array in an environment such
as polar ice, where the random noise level is low because the internally 
generated noise of the photomultiplier tube
is reduced at cold temperatures and the externally generated background
light from radioactive impurities is negligible.
The AMANDA collaboration agreed to join the Supernova
Early Alert Network \cite{kate} to 
confirm galactic supernova and determine the direction by triangulation
of the neutrino wavefront, which can precede the photon signal by several
hours or more.  The polar location of AMANDA simplifies the
task of triangulation, but the angular resolution
achieved by the SuperKamiokande experiment \cite{Beacom} may be 
superior. Neutrino observatories could search for
nearby extragalactic bursts by improving the collection area of the
optical sensors, implementing techniques to reduce the intrinsic noise,
and increasing the number of sensors in the array beyond several thousand.

 The recently reported evidence for
neutrino oscillation \cite{oscillation} in the atmospheric
neutrino data has triggered
the neutrino telescope community to investigate the physics capabilities
of their detectors for this particular science objective.
The energy spectrum of atmospheric neutrinos (hashed box, Fig.2 -- taken
from Protheroe's review paper (ref. 12)) is a
 steep power law, suggesting that the detected events will be predominantly
medium energy and the rate will be influenced by the energy threshold.  
Therefore, using atmospheric neutrinos to search
for neutrino oscillation requires energy thresholds of 5-20 GeV.
Detectors, such as Baikal 
NT-200 and NESTOR, or the insertion of high density strings into the 
AMANDA-II array, 
are designed to achieve this goal.  

\begin{figure}[h] 
\centering
\epsfxsize=8cm\epsffile{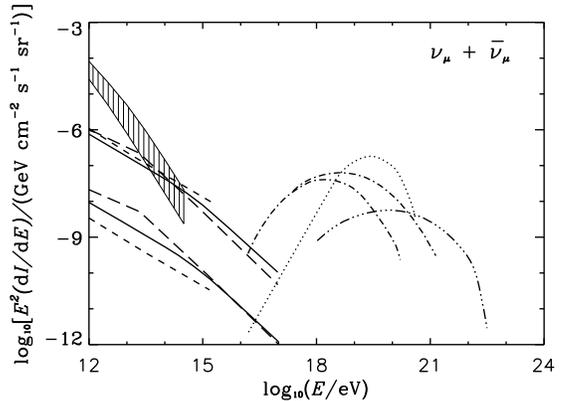}

\caption{Diffuse neutrinos from the atmosphere (hashed box), from GRB(thick
solid and dot-dash lines), from proton blazar models of AGN (thin solid,
 dashed, and  dotted lines). See Protheroe's review paper for a detailed 
explanation of the curves (ref 12).}
\end{figure}

Atmospheric neutrinos may reveal neutrino oscillations in several ways. 
A deviation from the expected angular distribution would be strong 
evidence for oscillations. Neutrino detectors can contribute to this science
by virtue of their large detection area  and consequent increase in 
statistical significance.  Unfortunately, these are difficult measurements
for neutrino arrays.  For the simplest case of two oscillating neutrino
species, the probability that a neutrino $\nu$ of flavor ${\it i} (e,\mu,\tau)$
will oscillate into a different flavor ${\it x}$ is given by

$P(\nu_{i}\rightarrow\nu_{x})= sin^{2}2\theta sin^{2}(1.27\Delta m^{2}
\frac{L(km)}{E_{\nu}(GeV)})$

where $\theta$ is the mixing angle, $\Delta m^{2}$ is the difference in mass
 squared in $eV^{2}$ of the two mass eigenstates,
L is pathlength between generating vertex and detector, and $E_{\nu}$
is the energy of the neutrino.

Unless the neutrino-induced muon event is completely contained within
the detector, the neutrino energy is not well measured. For the
current generation of neutrino detectors, through-going upward muons
are the most likely detection mode, but this only establishes a
lower limit on the neutrino energy.  Moreover, 
the energy threshold for muons which traverse the array is relatively high, so 
as $E_{\nu}$ increases, angular deviations become very subtle. For parameters
of $\Delta m^{2} = 2.5\times10^{-3} eV^{2}$ and maximal mixing, the
angular and energy dependence of the detector area must be
determined to 5\% or better.  It remains to be seen if this accuracy
can be achieved in practice.  Also $\nu_{e}$ events must be differentiated from
$\nu_{\mu}$ events. 

 A second idea takes advantage of the particular strengths of the existing 
neutrino arrays.  The linear symmetry of string-based designs results 
in excellent
sensitivity to nearly vertical tracks. The long lengths of instrumentation
can contain neutrino-induced events over a large interval of energies.  
By concentrating on nearly vertical tracks, backgrounds are easier to
reject.  The small vertical spacing of optical sensors (compared to
the horizontal spacing of the strings) reduce the energy threshold to
interesting levels.  The detection efficiency as a function of
energy can be calculated more accurately than for the entire hemisphere.  
In addition, the AMANDA array can calibrate its vertical sensitivity with
a well 
defined muon beam using coincidence events that simultaneously trigger 
another array at 900 meters.  If the 
vertex is contained within the central part of the array, then the 
light from the interaction vertex and outgoing muon can be modeled to
establish the energy of the neutrino with sufficient accuracy. Obviously,
the event rates are much lower for a restricted solid angle, but the
large detection area results in sufficient statistics.  However, the same
concern about being able to differentiate $\nu_{e}$ and 
$\nu_{\mu}$ events applies to this technique.  

A third method to search for neutrino oscillation over long pathlengths
(or baselines) utilize existing accelerators to direct a beam of 
$\nu_{\mu}$ particles with a known energy 
spectrum toward large neutrino telescopes located at distances
between 1000 and 10000 km.  While most discussion has involved
CERN and planned neutrino telescopes in the Mediterranean, the idea works 
the same for any accelerator and neutrino observatory as long as a 
neutrino beam can be pointed in the right direction. 

For kilometer-scale detectors, a significant fraction of neutrino-induced 
muons will be contained within the actively instrumented volume, so a
calorimetric measurement of the neutrino energy is possible. 
However, the larger spacing between sensors results in higher energy
thresholds which may be above the energies of interest. 
 Medium energy physics objectives can be retained 
if the kilometer-scale array surrounds a first generation neutrino array.  
The composite detector can identify and reject atmospheric muons ,
reducing background rejection requirements in the denser central region of 
the composite array.   

Neutrinos may be emitted from the center of the sun or earth as a consequence
of the annihilation of weakly-interacting cold dark matter particles (WIMPs)
that accumulate at the centers of these objects.  Galactic WIMPs, scattering
off nuclei, lose energy and may become gravitationally trapped. One interesting
class of WIMP candidates arise from minimal supersymmetric (SUSY) theory. 
Within this framework, Bergstrom et al.\cite{Bergstrom} have 
calculated the discovery potential for neutrino observatories and
beautifully illustrate their power to complement other 
search methods. Apparently, 
the parameterized ignorance of SUSY models is too vast to be completely
constrained by a single search technique.  Bergstrom et al. have attacked this
worrisome deficiency by combining the limits from cosmic ray antiproton 
instruments with the anticipated sensitivity of gamma ray satellites and 
neutrino observatories.  A comprehensive search strategy for SUSY particles
benefits enormously from the complementary information provided by
neutrino telescopes.  Combining astrophysical data from special purpose
and multipurpose survey instruments creates an intriguing blueprint for
future search strategies.

The primary motivation for very large neutrino telescopes is to identify
galactic or extragalactic sources, which may be point-like or diffuse.
The high energy frontier holds the most promise to achieve this 
scientific priority.  The atmospheric neutrino and muon backgrounds are 
lower, the effective area of the detector is larger, and angular resolution 
is likely to be better. Detection of diffuse sources requires good energy 
resolution with well understood tails but only marginal angular resolution. 

Theoretical activity has centered on modeling two classes of objects: galaxies
with active nuclei, or AGNs, and gamma ray bursts (GRBs).  These objects
are known to emit high energy photons, and may also be the accelerators of the
highest energy cosmic rays.  At TeV energies, the luminosities of some AGN 
are observed to flare by an order of magnitude in about a day, suggesting 
very compact central engines.  Models of the acceleration mechanism within
AGN differ ingeniously. The intensity of neutrino emission ranges from 
negligible in models that rely solely on electron acceleration to 
detectable in the most optimistic models based on hadron acceleration. 
Neutrino observatories are likely to play a key role in settling the debate.

If hadronic acceleration is present in AGN, then a diffuse
glow of neutrino emission should be observed uniformly over the sky, 
originating from distant (and more powerful) AGN.  Fig. 2 shows the 
energy spectrum for a representative sample of neutrino models.  

Figure 3, also taken from Protheroe,  converts the 
neutrino intensity  predictions into an event rate for a detector with
an effective area of 0.1 $km^{2}$. The calculations include absorption
by the earth, which becomes important for energies $\geq$100 TeV
\cite{Gaisser97}.
Diffuse sources could be distinguished from atmospheric neutrino background
by a flattening energy spectrum
above 100 TeV.  Some models can be differentiated by the their cutoff at
the highest energies and spectral shape.  Excellent energy resolution will
be necessary to select events with high energies and eliminate the lower 
energy atmospheric neutrino background. The representative models show
that there is little reduction in signal until the energy threshold
exceeds 10-100 TeV, with the exception of the atmospheric neutrino signal.
Theoretical considerations place a premium on detectors which attain
excellent performance at high energies.

\begin{figure}[h] 
\centering
\epsfxsize=8.0cm\epsffile{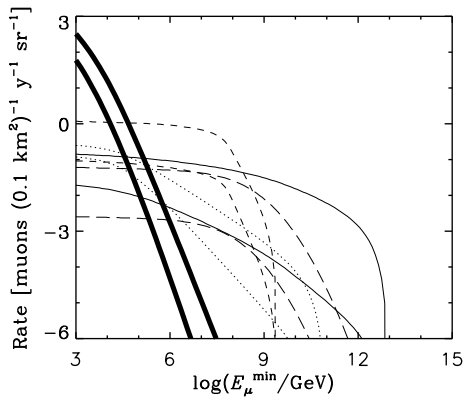}

\caption{ Rates for various neutrino sources.  Atmospheric neutrino-induced
muons (thick black lines); GRBs (dotted curves); $p\gamma$ proton blazar (shortdashed curves); topological defect model (thin solid curve), cosmic ray on
CMBR (long dashed curves).  Upper curves show horizontal signals and lower
curves show upward vertical signals.  For details, see Protheroe review.}
\end{figure}

A few caveats should be kept in mind when interpreting the previous figures.  
(1)The only ``background'' shown in Fig. 2 and 3 is atmospheric
neutrinos, but the rejection of down-going atmospheric muons represents a 
non-trivial hurdle that must be surmounted.   
(2)Point sources can be located to within a small fraction
of a steradian, and the atmospheric neutrino background decreases accordingly. 
Signal significance increases as $\sim \sqrt{A_{eff}}/\delta(\theta)$,
 where $\delta(\theta)$
is the angular extent of the source (or if considering a point source, 
proportional to the angular resolution of the detector.) 
(3)Correlated photon observations of GRBs by BATSE provide a 
special opportunity.  Events rates are determined by integrating over
all GRB events, and predicted to be $\sim\!50/$year \cite{WB97}.  However, the 
background livetime is only integrated over the duration of the bursts, 
which is $\sim 10^{-5}$ years.  In addition, the search for neutrino 
emission from GRBs is greatly simplified by the contemporaneous direction 
measurements by satellites.  Assuming a directional accuracy of 6 degrees,
the background is reduced by $d\Omega/(2\pi)$ = $5\times 10^{-3}$.
Combining directional and temporal information leads to a background 
reduction of 
$\sim 5\times 10^{-8}$ relative to a search for steady diffuse sources.
The relaxed rejection criteria increases the effective area of 
the detector, constrained primarily by the requirement to  maintain
sufficient angular resolution.  Alternatively, by raising the energy
threshold of the events, angular correlations may not be
necessary to reduce the background to manageable levels.  It is apparent that
searches for transient phenomena enjoy many experimental advantages due
to the reduced background (and consequent improvement in sensitivity).

Above $\sim\! 10 PeV$, the predicted event rates for optical
arrays with kilometer dimensions are too small.  Techniques based 
on detection of coherent
radio emission from neutrino-induced electromagnetic cascades are 
being pursued \cite{RICE}.  At radio wavelengths, the attenuation length
in ice is approximately 1 km, nearly an order of magnitude larger than
optical absorption lengths, suggesting that much larger volumes of ice
can be instrumented for a given number of sensors.
At the moment, more than a dozen radio receivers are buried in the same
holes used by the AMANDA collaboration to study reliability,  backgrounds,
calibration, and the feasibility of vertex reconstruction.
Long term issues such as power, signal transmission, servicing, and 
triggering over vast distances remain to be be solved.

Horizontal air shower techniques can be employed to explore the neutrino sky 
at extremely high energies \cite{Cronin}.  Conceivably, with
 $\sim\! 10 km^{3}$ of water equivalent target volume for $E_{\nu}>10^{19} eV$,
the Auger air shower array will have the sensitivity to search for 
neutrinos from cosmic ray interactions with the cosmic microwave background 
and more speculative signals from topological defects. 

Obviously, the desire to understand the optical and physical properties
of the local environment create many interdisciplinary opportunities.
Underwater neutrino observatories provide the facilities to monitor
the time variability of bioluminescence, temperature, salinity, water
currents, biofouling, etc.  The NESTOR collaboration has secured 
funding to deploy an optical cable from shore to the site off Pylos
instrumented with sensors of interest to oceanographers and neutrino
physicists \cite{LAERTIS}.  Multidisciplinary opportunities in
Antarctic ice led to the proposal to establish the DeepIce Science 
and Technology Center \cite{STC} (STC).  For example, DeepIce STC will promote 
interactions between seismologists and neutrino physicists to construct
a large seismic array for tomographic studies of the earths interior.
The Baikal detector monitors the seasonal water exchange processes in 
this unique Siberian lake \cite{Spiering98}.

\section{High Energy Neutrino Observatories}

The visionary decision by the DUMAND collaboration over 25 years ago 
to construct a large telescope nearly 5000m under the ocean and 40 km
from shore launched the experimental effort to construct a neutrino
observatory.  The design goals then were much the same as they are now:
threshold energy for neutrino detection $\sim\! 10-100 GeV$, effective
 detection
area = 20,000$m^{2}$, number of optical sensors = 200.  Unfortunately,
this pioneering effort fell victim to expensive logistical difficulties
and was de-funded.

At present, four groups are competing in the construction of 
high energy neutrino  observatories:  two in the Mediterranean -- NESTOR
 \cite{NESTOR,LAERTIS} 
and ANTARES\cite{ANTARES}-- one in Lake Baikal, Siberia, called NT-200 
\cite{Baikal} -- and one 
in deep ice at the South Pole called AMANDA \cite{barwick,B4,ICHEP98}. 
Baikal's NT-200 and AMANDA are
currently in operation, and feasibility studies are being carried out at
the Mediterranean sites.  The geographical
location is shown in Fig. 4.

\begin{figure}[h] 
\centering
\epsfxsize=8.5 cm\epsffile{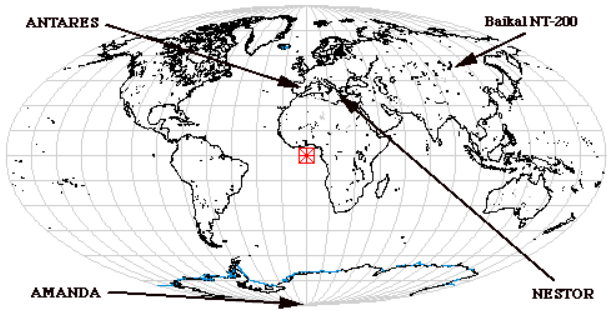}

\caption{Geographical location of operating or planned high energy
neutrino facilities.}
\end{figure}

AMANDA anchors the effort in the 
southern hemisphere and complements the sky coverage of the Siberian
and planned Mediterranean observatories.  Several new concepts for
surface neutrino observatories are being discussed \cite{HANUL}, but I 
will not cover those ideas here.

The Baikal collaboration has been accumulating experience with the 
construction and operation of water-based neutrino observatories since
1993, the longest track record of any group.  Those initial efforts
were followed by intermediate stages of construction that  include 
configurations with 96 and 144 optical sensors and culminate with
NT-200, which was completed in April 1998.  It consists of 192 optical
sensors positioned at a depth of 1.1 km below the surface of the lake.
The sensors are arranged in pairs and operated in coincidence to suppress 
unrelated signals from bioluminescence and internally generated random noise.
Deployment, the "Achilles Heel" of remotely located neutrino
observatories, has been solved by utilizing the seasonal ice cover on
Lake Baikal.  The solid platform can be accessed for significant
periods of time, enabling reliable detector assembly and repair of 
detector elements.

An umbrella-like frame maintains eight vertical strings of optical sensors,
consisting of a glass pressure vessel and a photomultiplier tube (PMT)
with a diameter of 37 cm.  The operation and performance of the Baikal 
detector is understood.  They have shown that the optical properties of
the water medium and 1.1 km depth are adequate to measure the angular
spectrum of atmospheric muons with good accuracy and to identify 
atmospheric neutrinos with the 96 element array (see Fig. 5).  This result 
bodes well for the Mediterranean sites because they are deeper and
their optical properties are better.   Neutrino events were extracted from
70 days of livetime.  After reconstruction, neutrino events were selected
by imposing a restriction on the  chi-square of the fit and requiring 
consistency between the reconstructed trajectory and the locations of
sensors registering photons.  In this context, sensors that do not
register photons carry important information as well.  Finally, the
non-gaussian tails of the angular distribution were reduced by 
imposing the condition that events must traverse more than 35m within
the array. 
 
\begin{figure}[h] 
\centering
\epsfxsize=8.0cm\epsffile{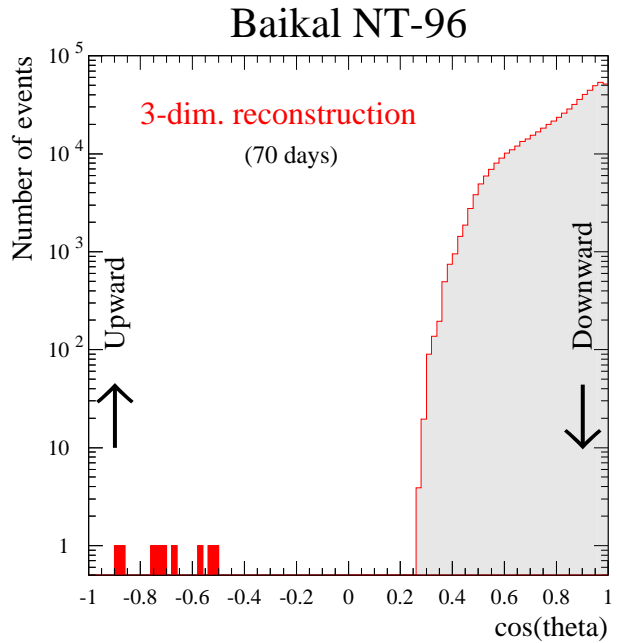}

\caption{Angular distribution of events in Baikal-96.}
\end{figure}

The high PMT density of the NT-200 design results in a low energy threshold
- advantageous for medium energy science goals - but limits the effective 
area at high energies to $ \sim 5\times 10^{3} m^{2}$, presumably too small to 
detect neutrinos from non-terrestrial sources.  A strawman design for
a 2000 sensor array has been presented.  The
effective area would be $\sim\! 10^5 m^2$, while retaining a 10-20 GeV
energy threshold. It could fill the niche between current generation of
neutrino detectors and future kilometer-scale arrays with, presumably,
much higher energy thresholds.

A flurry of research and development activities have occupied the 
NESTOR and ANTARES collaborations as they assess the relevant physical
and optical parameters of their sites.  Deployment methods are 
being developed and refined through a series of operations using
barges, research and military vessels.  The NESTOR and ANTARES groups 
envision quite different deployment schemes, array designs, and signal
processing. Technological solutions are being sought which are 
affordable, reliable, and expandable.

Over the past few years, the ANTARES collaboration has methodically 
determined the critical 
optical parameters of a 2400m deep site off the coast of Toulon, France.
Significant R\&D has concentrated on string deployment and retrieval.
They have reported that one string has been installed
at the site and recovered after one year of flawless operation.  
This success paves the way for more complex and difficult operations, such 
as the deployment of a fully functional string of sensors, deployment of 
multiple strings, or the insertion of a string within an existing array.

  Precision attenuation (Fig. 6) and scattering measurements at 
wavelength of 450nm   
are extremely encouraging.  The deliberate development plan calls for
the construction of a demonstration array, consisting of 100-200 optical
sensors, by the end of 1999 (although the configuration and schedule are
subject to change).

\begin{figure}[t] 
\centering
\epsfxsize=8cm\epsffile{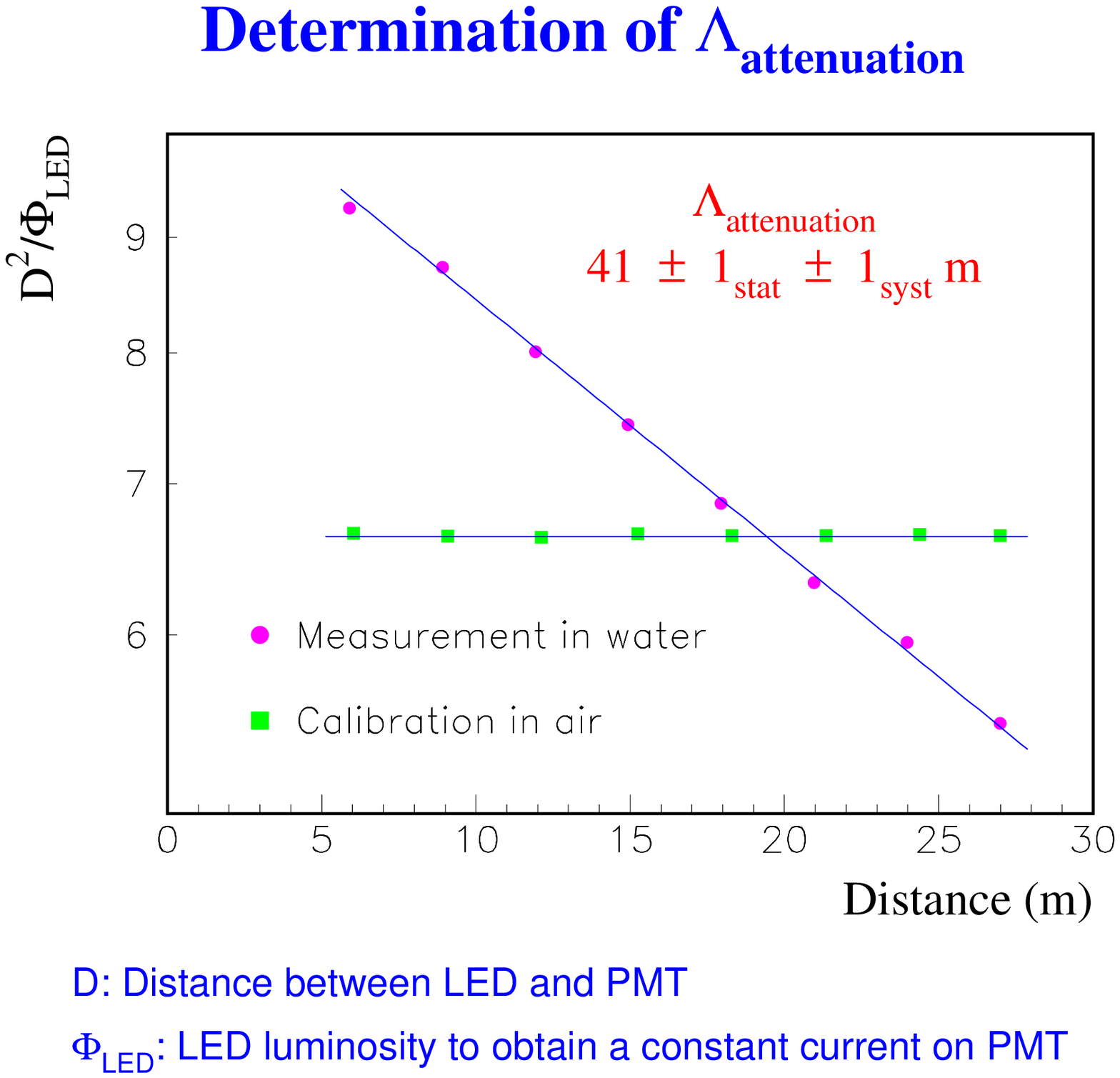}

\caption{Attenuation length for water at ANTARES site (at 450nm).}
\end{figure}

Environmental studies at the Toulon site show that upward facing PMTs 
lose sensitivity over time due to the accumulation of organic debris,
so the ANTARES design consists of only downward looking PMTs.  Deep
sea currents have been measured over a period of a year and show no
unusual excursions from expected values.  

Simulations of an array consisting of 15 triads of strings ($\sim\! 1000$ PMTs)
indicate that neutrino events can be cleanly identified.   Random noise 
exceeding 50 kHz per optical sensor has been measured, but
can be eliminated by straightforward coincidence requirements between
neighboring elements in the array. Bioluminescent flashes do not affect 
local coincidence rates due to the relatively weak intensity of the
output and the relatively long duration of the burst. Muon directions
should be identified with sub-degree angular resolution.

NESTOR plans to deploy an array of 168 optical sensors at a depth
of 3.5-4.0 km of the cost of Pylos, Greece.  The large depth significantly
reduces the background of down-going atmospheric muons, but places greater
stress on the penetrator connections.  Hexagonal floors, rather
than strings, comprise the basic unit.  The array consists of 12 floors,
fixed in place with an extensive network of wire guides, and assembled 
to form a 
200m tall tower.  Site testing is complete, showing excellent optical
properties.  Like the Baikal design, a symmetric up-down arrangement
of PMT orientations will insure better uniformity in its angular
acceptance.  Upward facing PMTs are thought not suffer from obscuration
due to sedimentation or biological growth.  The array design is 
expected to achieve
low energy threshold due to the relatively high density of optical
sensors.  Horizontal separations between optical modules on a given floor
are slightly larger than 30 meters.
Recently, the NESTOR collaboration has performed mechanical tests by 
successfully towing a single floor out to sea and deploying to a 
depth of ~2600 m.  In the near future, a far more ambitious plan to
deploy two, fully instrumented, floors to depth.  It is hoped that
these tests will establish the electro-mechanical durability of the 
signal processing and transmission systems.

AMANDA was completed in early 1997 and has operated reliably since. The
rapid growth of the AMANDA program makes it a strong candidate for 
expansion to kilometer scales.
From its inception, AMANDA has been designed to confront the robustness
issue, which has been the primary technological difficulty of water based
arrays.  Three months of access to a solid platform insures that sufficient
time is available for certification and quality assurance assessment.
The infrastructure at the South Pole, soon to be expanded and modernized,
provides a reliable transportation system, communications, adequate power, 
laboratory facilities, and the resources to solve unexpected difficulties in
real time.  The distance between the optical sensors and the surface 
facilities is only a few kilometers, so a highly redundant array 
architecture could be implemented with no potential single-point catastrophes. 
It also accommodates a modest failure rate of individual sensors.  The the 
architecture permits both analog and digital signal processing solutions.
A conservative approach to the {\it in situ} hardware was adopted.  Given
the geographic novelty of the South Pole site,
a simple, but mechanically and electrically robust design based on
analog transmission through copper cables was implemented. Experience
from time-of-flight detectors indicated that adequate timing
resolution could be achieved despite the signal dispersion in the long
cables. High gain PMTs were developed to compensate for the large cable
dispersion and attenuation. The extraordinary thermal stability of the ice
 suggested a very low
rate of drift in the calibrated parameters, which in turn simplified 
manpower requirements.   
Unlike water, ice was recognized to present few long term electrical problems
and durability of the PMTs benefitted from the stable, sub-zero, 
ambient temperatures.

Several unusual obstacles confronted the AMANDA program.  The architecture had
to accommodate a large uncertainty in the optical properties of polar ice,
which were poorly known prior to the initiation of the AMANDA campaign, 
and in hindsight, overly naive.  Nevertheless, the AMANDA
collaboration has shown that Antarctic ice is a suitable medium for
a neutrino observatory.  Remarkably, absorption in Antarctic ice
is far better than expected from laboratory measurements. However, the 
transition depth to bubble-free ice at the South Pole was initially 
under-estimated \cite{Gow}, and the magnitude of the residual 
scattering in the bubble-free region had to be determined experimentally.
Models of the ice structure were greatly improved with the initial data from
800-1000m \cite{Price_ice}.  While the transition to 
bubble-free ice was never in doubt - it being firmly established
by deep ice-cores from Greenland and Antarctica-  the
depth of the transition was reliably bounded by the new models.  With
the deployment of the first four strings of AMANDA-B, the bubble-free 
transition was confirmed and fell within the predicted range. Models
of the ice could also bound absorption and scattering lengths, but
the uncertainties in the model parameters were large.
By measuring the the timing distribution of pulses of laser
light as a function of wavelength and depth \cite{ice_prop} the optical
properties were measured with the the necessary precision. The
experiments required that the geometry of the array, timing parameters, 
and optical properties had to be measured in situ.  To complicate matters,  
the determination of the geometry and optical properties were inter-dependent,
so an iterative method of data analysis was developed to address this
feature.  Fortunately, once the geometry and ice properties are known, 
they do not change.  

Throughout the development of AMANDA, riskier but more capable technologies
have been investigated.  A rigorous, deliberate evaluation process was 
instituted which required that laboratory prototypes be installed in-situ and
integrated into the existing array to assess reliability, deployment,
logistic, and system capatibility issues.  Old and new technologies 
were combined into hybrid modules, retaining the reliability of the
previous methods while evaluating the newer ones.  Reliable baseline 
technologies have been phased out as confidence in the new technologies grew.
Both analog and digital technologies are realistic options for
signal transmission.  Analog technologies, based on laser diode transmitters
and optical fibers, offer many advances over the current baseline, including
excellent signal fidelity, improved dynamic range, low cross-talk, simplified
calibration and debugging.  Digital solutions are being explored as well.  
Full waveform digitization 
within the module insures that the event contains maximum information.
Multiplexing of several sensors on a single cable, not easily implemented 
in analog transmission, can 
reduce costs assuming the reliability justifies this cost reduction.  Fiber
optic cables, one of the big ticket items in the AMANDA-II design, can be 
eliminated in the digital architecture.

The status of the AMANDA project can be summarized as follows:

\begin{itemize}

\item Construction of the first generation AMANDA detector \cite{barwick}
was completed in the austral summer 96--97. It consists of 302 optical modules,
located on 10 separate strings, deployed to depths between 
1500--2000~m; see Fig.\,7. An optical module (OM)
consists of an 8\,inch photomultiplier tube (R5912-02) encapsulated in a 
glass pressure sphere and mounting hardware. Analog signals are sent to
the surface via electrical cables in AMANDA-B10. The conservative
design has resulted in an {\it in-situ} failure rate of only 3 \%.

\item Data taken with the first 4 strings (a total of 80 OM's), deployed in 
January of 1996 to assess
the optical properties of the deep ice, have been analyzed. This partial
detector will be referred to as AMANDA-B4. Nearly vertical up-going muons 
are found at a rate that is statistically consistent with the expected flux 
of atmospheric neutrinos. As Fig. 8 shows, events are clearly 
separated from the background population \cite{B4} of poorly reconstructed
downgoing muons.  
Simulations and data agree - from a crude 
check of hardware trigger rate to careful examination of the 
muon angular distributions as the selection criteria are refined.  Starting
with $10^5$ events which pass the trigger, a set of selection criteria were
sequentially implemented to reduce the number of events misreconstructed
as upward going.   Selection criteria were improved until no events remain 
in the up-going direction.  Absolute events rates agree to within a factor of 
3 at all stages of this analysis, limiting the absolute error in the Monte
Carlo estimates of effective area.

\item The commissioning phase of the full detector is now completed (July '98)
and analysis of data from 1997 is in progress. Final calibration of array 
geometry, cable-dependent time delays, and PMT performance was completed after
the return of the first year of full operation.  First-look analysis indicates 
that events can be extracted with
trajectories in the upward direction. A more extensive evaluation of 
background and detector performance is currently in progress.

\end{itemize}

\begin{figure*} 
\centering
\epsfxsize=15cm\epsffile{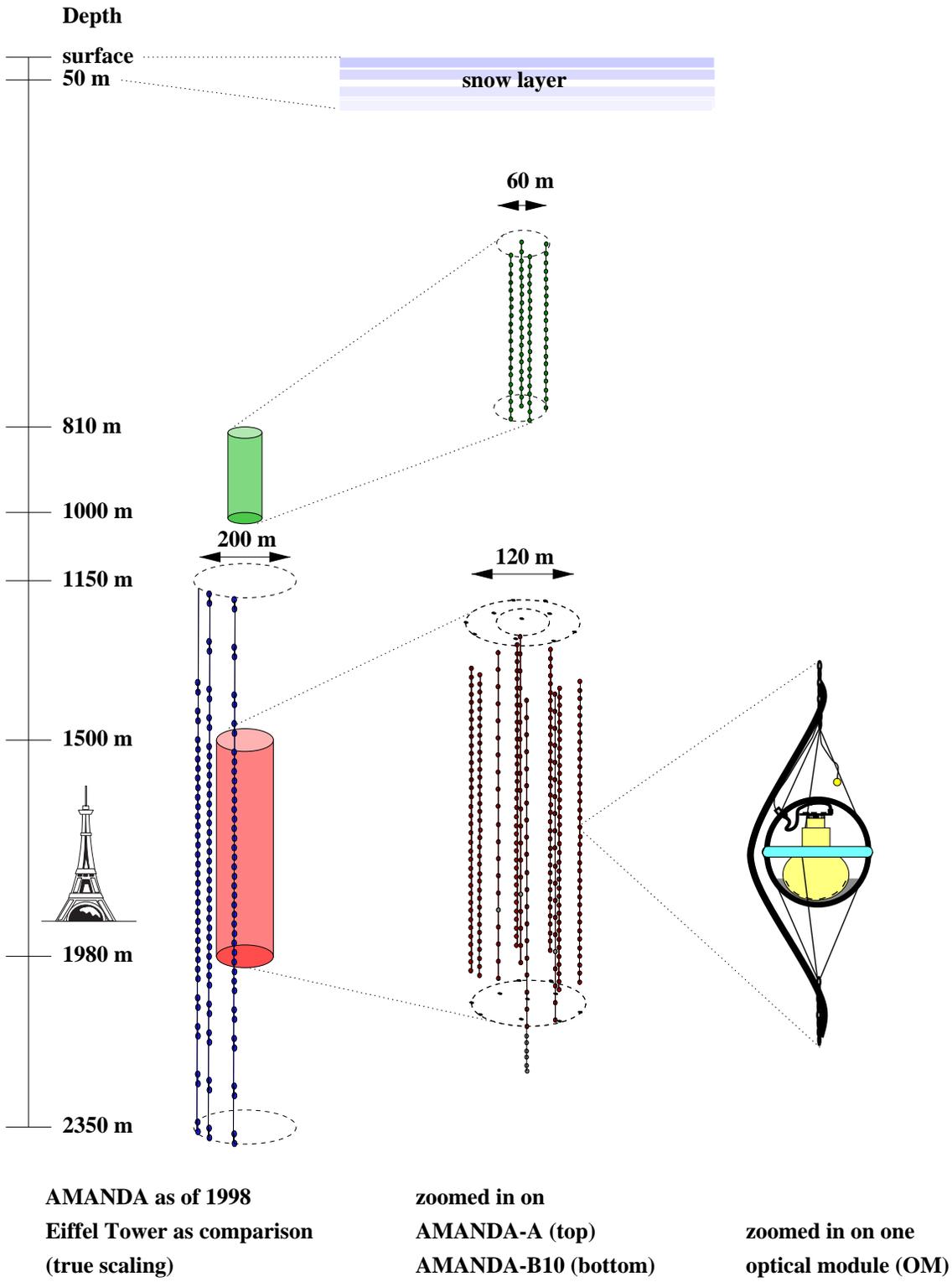}

\caption{Configuration of Antarctic Muon And Neutrino Detector Array (AMANDA)
 in 1998.}
\end{figure*}

AMANDA-II is an approved and funded expansion of the AMANDA-B array.  The
proposed array consists of 11 additional strings of OMs arranged concentrically
around AMANDA-B10.
Current simulations predict that AMANDA-II will have an effective 
detection area of
$\sim\! 3\times10^4 $m$^2$ (depending on energy; significantly less for 
atmospheric neutrinos and somewhat larger for PeV-scale neutrinos) and 
angular resolution of $\sim\! 1^\circ$ (again, depending slightly on energy).
Construction of the AMANDA-II upgrade began in January 1998 with the deployment
of three strings to a depth of 2350m.  Each string contained 42 OMs that
were positioned along the lowest kilometer of cable.  Thus, these strings 
serve as a full-scale prototype for
a planned expansion to a kilometer-scale array of sensors called IceCube.

The deployment of AMANDA-II strings in 1998 addressed both science
and R\&D goals.
First, the optical properties of the ice at depths above and below AMANDA-B10
were measured.  These results will be used to optimize the depth and spacing
of the remaining eight strings of AMANDA-II sensors. Second, the longer 
lever arms of the new strings provides crucial data to verify simulation
results on event topologies not readily obtained by AMANDA-B10.

\begin{figure}[h] 
\centering
\epsfxsize=8cm\epsffile{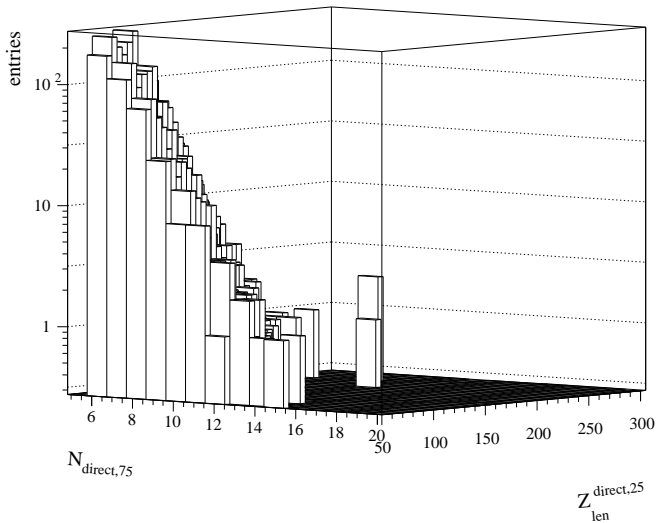}

\caption{Three high energy neutrino candidates are resolved from the
population of background. Shown are two variables from analysis of data
from the first 4 strings in AMANDA. $N_{direct}$ counts the number of 
modules that registered photons within 75ns of the predicted time and
 $Z_{len}$ specifies the maximum distance between sensors with direct
hits.}
\end{figure}

A pair of TV cameras were lowered into the last hole.
The resulting images visually confirm the exceptional clarity of the ice 
deduced from calibration measurements.  The fidelity of signal transmission 
was dramatically improved by transmitting analog signals from the PMT to
the surface over optical fiber and electrical cable simultaneously.  Optical
transmission of signals, using an LED, eliminates the distortion of the PMT
waveform while preserving many aspects of the conservative design features 
introduced by analog signal transmission over electrical cable.  The high 
fidelity of signal reproduction at the surface improves the
double-pulse resolution by an order of magnitude.  Reconstruction should
benefit from better identification of multiphoton signals, and from 
reduced cross-talk.  Time-delay calibration procedures are simplified so 
fewer manpower resources are required.
  
The robustness of the optical fiber cables and connectors was improved.  One 
combination of fiber and connector technologies produced a 90\% survival rate.
Based on the success of the hybrid optical technologies, the remaining 
AMANDA-II OMs will transmit analog optical and electrical signals to the
surface.

Data from the first phase of construction, AMANDA-B4, has been used to 
measure the optical quality of ice, geometric spacing {\it in situ}, and
angular distributions of atmospheric muons.
During the six month commissioning phase following the first year of
AMANDA-B10 operation, system calibration of the array geometry, propagation
constants, and gain drifts was completed.  Concurrently,
software was developed to reduce the 0.5 TB of data by a factor of 10
by filtering events that were readily identified as atmospheric muons.

With logistics, deployment, calibration, and durability issues solved,  
the AMANDA collaboration focussed its efforts on data analysis by
concentrating on the known atmospheric neutrino signal. Although the current
AMANDA design is optimized to search for higher
energy neutrinos, a sufficient number of atmospheric neutrinos can
be observed to verify critical performance parameters at medium energy.
The collaboration is developing suitable estimates of energy  for both
the muon and cascade events
. The measured energy spectra can help
to distinguish between many potential sources of high energy neutrinos.

Initial analyses have produced
upward-going muon events which possess several important topological 
features that are consistent with atmospheric neutrinos and inconsistent
with background.  Fig. 9 shows one example of this type of event. Optical
modules which detect Cherenkov light are color coded according to the 
rainbow with 
the earliest hits in red and latest in blue.  The topology of the detected
light is consistent with an upward traveling muon, and inconsistent with
timing and topological characteristics of $\sim\! 20$ million simulated 
background events. 

\begin{figure*} 
\centering
\epsfxsize=15cm\epsffile{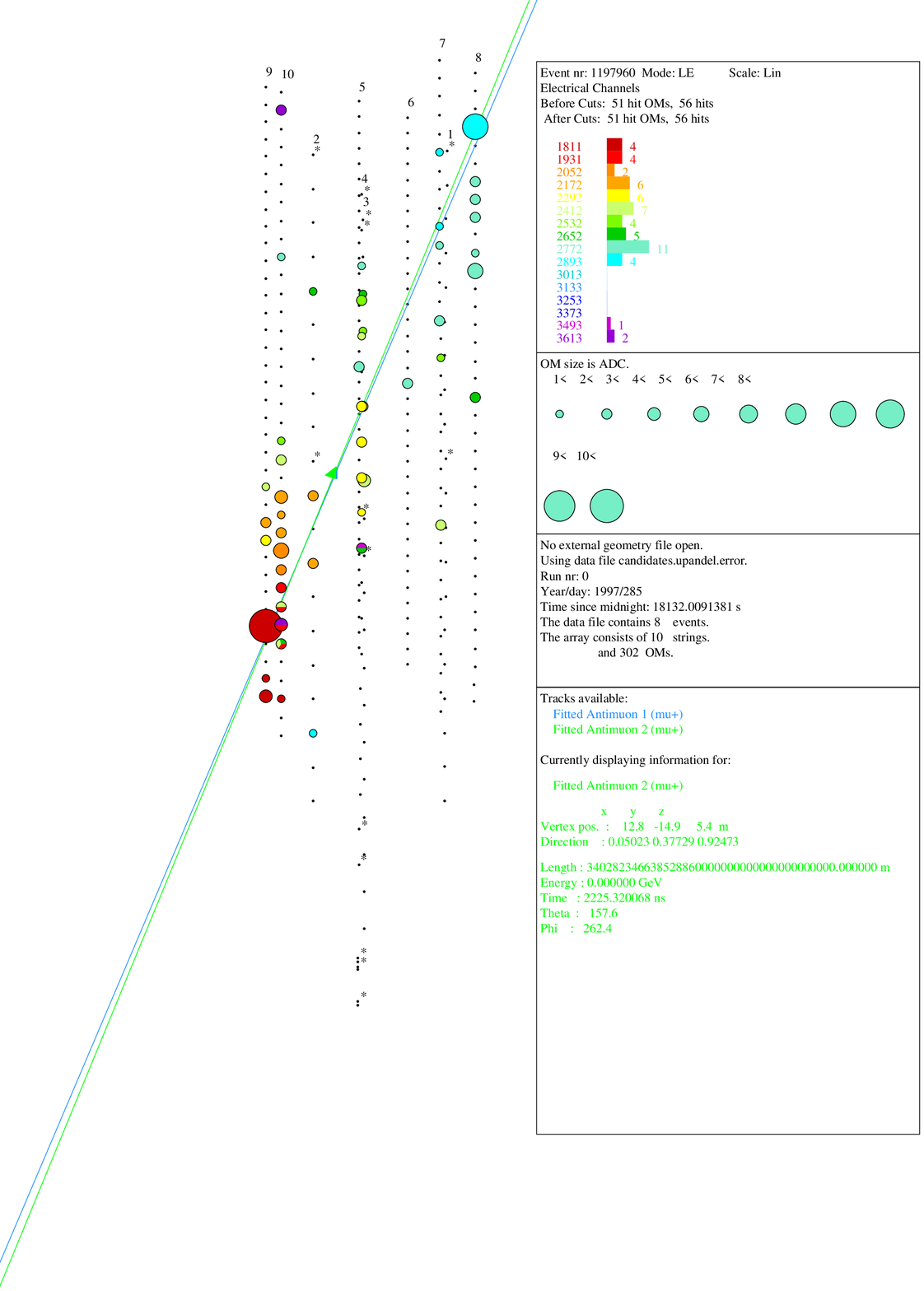}

\caption{Example of event which is reconstructed to have an upward
going trajectory in AMANDA-B10. Circle sizes are proportional to 
signal amplitudes.}
\end{figure*}

At this early stage of analysis, the reconstruction techniques and selection 
criteria are continuously being refined so improvements are expected.  
While AMANDA is capable of detecting a
sufficient number of atmospheric neutrinos to verify detector operation,
the event rates are low due to the soft energy energy spectrum and
the strong background rejection criteria required to identify a diffuse
source.  As more data is analyzed and efficiencies improve, 
the comparison of the angular distributions generated by simulation and 
data will provide a stringent test of the atmospheric neutrino hypothesis.

Searches for steady-state point sources require selection 
criteria that are optimized for high energy signals.  At $\geq$TeV
energies, background  rejection is aided by angular correlations with 
known gamma (GeV-TeV) sources.  In addition, larger energy thresholds can be 
imposed during the analysis because the average light output of the
muon above $\sim\! 1 TeV$ depends linearly on the energy, although the 
output is highly stochastic which limits the energy resolution. This
situation will improve as the neutrino facilities increase in scale
and become more symmetric.

\section{Future Arrays with Kilometer Dimensions}

It may not be strictly accidental that deployment methods based on
solid surfaces have enjoyed greater success at this moment, but there
is little doubt that deep water deployment can be done.  In the long
term, they will be challenged to demonstrate that reliability and 
cost issues remain competative with AMANDA and NT-200.  
While the current generation of  neutrino observatories 
represent remarkable achievements, they are only a fraction of the
size ultimately required to probe the hadronic sky. In
fact, all current programs have the potential for expansion to kilometer
scales  - it is one of the important design requirements of the current
generation of neutrino detectors.  Several arguments
have been used to coalesce around a detector with kilometer dimensions.  
A survey of theoretical models of GRB and AGN emission produce general
agreement at fluxes that would be detectable with kilometer scale detectors-
with orders of magnitude bracketing the maximum and minimum fluxes.
Given the current state of theoretical uncertainty, the bigger the
detector, the better the chances. More persuasively, the symmetric shape and
larger volumes offer significant experimental advantages:  particle 
trajectories are reconstructed with much higher efficiency, down-going
atmospheric muon background will be simpler to reject, and energy 
resolution will be improved, perhaps dramatically.  It
may be possible to distinguish each of the three known neutrino flavors 
\cite{IceCube}.

  Several workshops have been held worldwide to discuss ideas for future 
expansion of the neutrino observatories.  At UC-Irvine, for example,
 a workshop was held in
March 1998 to initiate the conceptual design of the IceCube Neutrino 
Facility in Antarctica.
Scientific goals and priorities were actively debated, and the sensitivity of
several strawman designs were studied within the rough constraints of 
5000 OMs and fewer than 80 strings.  A reasonable estimate of cost, scaling 
from the default analog-based technology, is \$7000 per optical sensor. 
Deployment and logistics costs must also be accounted for.  The construction 
of IceCube may be completed in 5 years given a reasonable projection of the 
drilling capacity. The Baikal collaboration envisions an expansion to 2000 OMs.
Similarly, the NESTOR and ANTARES groups anticipate significant expansion 
after successful operation of the first generation detectors.

\section{Conclusions}

The late Fred Reines, Nobel Laureate and father of neutrino
physics, was fond of saying that one should choose to work on physics
topics worthy of a lifetime's study.  The broad diversity of scientific
capabilities and enormous potential of
high energy neutrino astrophysics certainly qualifies. In view of the large
number of possible sources discussed by theorists and even larger variation in
their predicted intensity of neutrino emission, it is plausible that some
will be detected by current, or soon-to-be upgraded, neutrino detectors such
as AMANDA-II.  If history is a guide, there will be surprises as well as
these detectors begin to survey the great canvas of the unknown.

High energy neutrino facilities are developing during an era of exciting
discoveries in related areas of particle astrophysics:  the detection
of rapidly varying multi-TeV gamma ray signals from AGN, the discovery that
GRBs are extremely distant, the reports of cosmic rays exceeding $10^{20} eV$
-beyond the Greisen-Zatsepin-Kuzmin limit, and strong evidence for neutrino
oscillation from atmospheric neutrino data.   
At the close of millenium, the hadronic sky is being probed with first
generation neutrino detectors.  They constitute bold, but essential, 
first steps toward the realization of multi-messenger astronomy. However,
much larger
facilities of kilometer-scale dimensions are required to examine the
sky at sensitivities beginning to approach scientific consensus.
Given suitable instrumentation, it is not unreasonable to imagine that 
the insights revealed by the neutrino messenger of the hadronic sky will 
soon rival those deduced by observing the electromagnetic sky.  
This is the challenge for the next millenium.

\section*{Acknowledgements}

The organizing committee deserves special commendation for their professional
and cheerful management of this symposium. I would like to thank my AMANDA
collaborators for their support and interest.   S.W.B
~is supported in part by the University of California--Irvine and in part by
NSF grants OPP-9512196 and PHY-9722641.


\section*{References}
\begin{thebibliography}{99}
\frenchspacing

\bibitem{pr} For a review, see T.\,K.\,Gaisser, F.\,Halzen and T.\,Stanev,
{\it Phys.\ Rep.} {\bf 258}(3), 173 (1995); R.\,Gandhi, C.\,Quigg,
M.\,H.\,Reno and I.\,Sarcevic, {\it Astropart. Phys.}, {\bf 5}, 81 (1996).


\bibitem{barwick} S.\,W.\,Barwick {\it et al.}, {\it The status of the AMANDA
high-energy neutrino detector}, in Proceedings of the 25th International
Cosmic Ray Conference, Durban, South Africa (1997);  see also 
http://amanda.berkeley.edu/ .








\bibitem{berez77} V.S. Berezinskii and G.T. Zatsepin, {\it Sov. Phys.Usp.},
{\bf 20}(1977)361.

\bibitem{Gaisser97} T.K. Gaisser, OECD Megascience Forum Workshop, Taormina,
Sicily (May 1997),astro-ph/9707283.

\bibitem{B4} P. Askebjer, {\it et al.}, submitted to Astropart. Phys.

\bibitem{ICHEP98} S.W. Barwick,{\it et al.}, in Proceedings of the 
{\it International Conference on High Energy Physics(ICHEP98)}, Vancouver,
Canada, July, 1998.

\bibitem{TASI98} F. Halzen, Lectures presented at TASI School, July 1998,
astro-ph/9810368.

\bibitem{WB97} E. Waxman and J. Bahcall, Phys. Rev. Lett. {\bf 78},(1997)2292.

\bibitem{WB99} E. Waxman and J. Bahcall, Phys Rev. {\bf D59} (1999)023002.

\bibitem{Mannheim99}K. Mannheim, R.J. Protheroe, and J.P. Rachen, submitted
to Phys. Rev. D(1999), astro-ph/9812398.

\bibitem{history60} K. Greisen, Ann. Rev. Nucl. Science,{\bf 10}(1960)63;
F. Reines, Ann. Rev. Nucl. Science, {\bf 10}(1960)1; M. A. Markov and I.M.
Zheleznykh, Nucl. Phys. {\bf 27}(1961)385; M.A. Markov in{\it Proceedings of
the International Conference on High Energy Physics at Rochester}, E.C.G.
Sudarshan, J.H. Tinlot, and A.C. Melissinos, Eds.(1960)578; Proceedings 
of the DUMAND Summer Workshop, Bellingham, Washington (1975);  A. Roberts,
Rev. Mod. Phys. {\bf 64}(1992)259.


\bibitem{Proth99} R.J. Protheroe, in Proceedings of 18th International
Conference on Neutrino Physics and Astrophysics(Neutrino 98), Takayama,
 Japan(1998); astro-ph/9809144.

\bibitem{icehistory} F.Halzen and J. Learned, in {\it Proceedings of the
International Symposium on Very High Energy Cosmic Ray Interactions}, 
University of Lodz Publishers, M. Giler, ed,(1989); F. Halzen, J. Learned,
and T. Stanev, AIP Conference Proceedings {\bf 198}(1989)39.




\bibitem{HANUL} D. Normile, Science,{\bf 279}(1998)802.

\bibitem{ANTARES98} F. Hubaut, Proceedings of the Neutrino Workshop, Zeuthen,
Germany (July,1998, C. Spiering, ed).

\bibitem{Porrata} R. Porrata {\it et al.} in Proceedings of the 25th 
International Cosmic Ray Conference, Durban, South Africa {\bf 7} (1997)9;
R. Porrata,{\it The Energy Spectrum of Pointlike Events in AMANDA-A},
dissertation, Univ. of California-Irvine (1998).

\bibitem{kate} For more information, see 
http://www-ussk.icrr.u-tokyo.ac.jp/~kate/snnet/snnet.html

\bibitem{Beacom} J.F. Beacom and P. Vogel, astro-ph/9811350.

\bibitem{oscillation} Y. Fukuda {\it et al.}, Phys. Rev. Lett.{\bf 81}
(1998)1562; Galagher {\it et al.}, in Proceedings fo the 29th Intl. Conf.
on High Energy Physics (Vancouver, July, 1998); Michael {\it et al.},in
 Proceedings fo the 29th Intl. Conf.on High Energy Physics (Vancouver, 
July, 1998).

\bibitem{Bergstrom} L. Bergstrom, J. Edsjo, and P. Gondolo, Phys.Rev. D{\bf 58}
 (1998)103519 and hep-ph/9806293. 

\bibitem{RICE} C. Allen, {\it et al.}, in Proceedings of the 25th 
International Cosmic Ray Conference,Durban, South Africa {\bf 7} (1997)85;
P.B. Price, Astropart. Phys.{\bf 5}(1996)43.

\bibitem{Cronin} J. Cronin, Nobel Symposium on Particle Physics and the 
Universe (Haga Slott, Sweden, 1998); see also
 http://www.fnal.gov/pub/pierreauger.html for more information.

\bibitem{LAERTIS} A. Capone, in Proceedings of the 25th 
International Cosmic Ray Conference,Durban, South Africa {\bf 7} (1997)49;
S. Sotiriou,  Proceedings of the Neutrino Workshop, Zeuthen,
Germany (July,1998, C. Spiering, ed). 

\bibitem{STC}P.B. Price, proposal to the National Science Foundation, USA
(1998)

\bibitem{Spiering98} C.Spiering, Proceedings of the Ringberg Neutrino 
Workshop, Germany (1998);  C. Spiering, Prog.Part.Nucl.Phys. {\bf 40}(1998)391.
 
\bibitem{NESTOR} L. Trascatti, Nucl. Phys. {\bf B70} (Proc. Suppl.)(1998)442.

\bibitem{ANTARES} F. Feinstein, in Proceedings of the Neutrino Workshop,
Zeuthen, Germany (July,1998, C. Spiering, ed); F. Feinstein, Nucl. Phys.
{\bf B70} (Proc. Suppl.)(1998)445; see also 
http://antares.in2p3.fr/antares/antares.html

\bibitem{Baikal} I.A. Belolaptikov {\it et al.}, Astropart. Phys. {\bf 7}
(1997)263; L. Kuzmichev, in Proceedings of the 25th 
International Cosmic Ray Conference, Durban, South Africa {\bf 7} (1997)21;
S. Klimushin,  Proceedings of the Neutrino Workshop, Zeuthen,
Germany (July,1998, C. Spiering, ed); G.V. Domogatsky, Nucl. Phys. {\bf B70}
(Proc. Suppl.,1998)439;
see also http://www.ifh.de/baikal/baikalhome.html


\bibitem{Gow} A.J. Gow and T. Williamson, {\it Cold Regions Research and
Engineering Lab, Research Report} {\bf 339}(1975).

\bibitem{Price_ice} P.B Price {\it et al.}, J. Glac.{\bf 41}(1995)445; P.B.
Appl. Opt., vol.36( 1997)4181;   Y.D. He and P.B. Price, J. of
Geophys. Res., vol.103(1998)17041.

\bibitem{ice_prop}  P. Askebjer {\it et al.} Science {\bf 267}(1995)1147; 
 P. Askebjer {\it et al.}Geophys. Res. Lett. {\bf 24} (1997)1355.

\bibitem{IceCube} J.G. Learned and S.Pakvasa, Astropart. Phys. {\bf 3}(1995)
267.; F. Halzen and D. Saltzberg, Phys. Rev. Lett., {\bf 81}(1998)430; 
 F. Halzen, Proc. of Snowmass 94, (R. Kolb and R. Peccei, eds.) 

\end{thebibliography}
\end{document}